\documentclass[aps,superscriptaddress]{revtex4}
\usepackage{amssymb,amsmath,epsfig}
\usepackage[colorlinks=true, pdfstartview=FitV, linkcolor=blue, citecolor=red, urlcolor=magenta, breaklinks=true]{hyperref}
\begin{document}
\title{The entropy of an acoustic black hole in neo-Newtonian theory}
\author{M. A. Anacleto}
\email{ anacleto@df.ufcg.edu.br}
\affiliation{Departamento de F\'{\i}sica, Universidade Federal de Campina Grande
Caixa Postal 10071, 58429-900 Campina Grande, Para\'{\i}ba, Brazil}
\author{Ines. G. Salako}
\email{ ines.salako@imsp-uac.org; inessalako@gmail.com}
\affiliation{Ecole de Machinisme Agricole et de Construction M\'ecanique -
Universit\'e d'Agriculture de    K\'etou,  BP:43 K\'etou, B\'enin}
\affiliation{Institut de Math\'ematiques et de Sciences Physiques
(IMSP) 01 BP 613 Porto-Novo, B\'enin}
\author{F. A. Brito}
\email{fabrito@df.ufcg.edu.br}
\affiliation{Departamento de F\'{\i}sica, Universidade Federal de Campina Grande
Caixa Postal 10071, 58429-900 Campina Grande, Para\'{\i}ba, Brazil}
\affiliation{Departamento de F\'isica, Universidade Federal da Para\'iba, Caixa Postal 5008, 58051-970 Jo\~ao Pessoa, Para\'iba, Brazil}
\author{E. Passos}
\email{passos@df.ufcg.edu.br}
\affiliation{Departamento de F\'{\i}sica, Universidade Federal de Campina Grande
Caixa Postal 10071, 58429-900 Campina Grande, Para\'{\i}ba, Brazil}
\affiliation{ Instituto de F\'{\i}sica, Universidade Federal do Rio de Janeiro,
Caixa Postal 21945, Rio de Janeiro, Brazil}

\begin{abstract} 
In this paper we consider the metric of a 2+1-dimensional rotating acoustic black hole in the neo-Newtonian theory to compute the Hawking temperature and applying the quantum statistical method, we calculate the statistical entropy using a corrected state density due to the generalized uncertainty principle (GUP). In our calculations we have shown that the obtained entropy is finite and correction terms are generated. Moreover, the computation of the entropy for this method does not present logarithmic corrections.
\end{abstract}

\maketitle
\pretolerance10000

\section{Introduction}
The study to understand the entropy of black holes is one of the most important issues in theoretical physics.
It has been proposed by Bekenstein and Hawking that the black hole entropy is proportional to its horizon area~\cite{Hawking,SWH,JDB,JDB85}. 
Since then, various methods have been proposed in the literature to explore the statistical origin of black holes entropy. {These methods are grounded on the understanding that 
the black hole entropy can be viewed as the entanglement entropy related to the degree of entanglement
between the modes in both sides of the horizon. In other words, the Bekenstein-Hawking entropy can be obtained via the statistical mechanics of a system around the horizon \cite{SWU,G-H}}.
One of such methods is the so-called brick-wall method introduced by G. 't Hooft~\cite{thooft}. To calculate the entropy by this method it is necessary to introduce an ultraviolet cut-off in order to eliminate the divergences in the density of states near the horizon of the black hole. In Ref.~\cite{Rinaldi,Rinaldi:2013aa,SG2011}, by applying the brick-wall method, one has been computed  {the rate of entropy production in the emitted Hawking radiation }  of an acoustic black hole in (1 + 1) - dimensions.

{ On the other hand,  as shown by the authors in~\cite{Brustein,Kim2006,Park2007,Sun2004,Yoon2007}, the brick-wall model does not present divergences when the GUP is considered.}
{In addition, the authors in~\cite{XLi,zhao2004,RZ2003,WK2006,KKP2007,JHa2007,YWKim2007} considering a density of states modified by the GUP computed the statistical entropy of many black holes.}
Moreover, considering the effects of the GUP in the tunneling formalism, the quantum-corrected Hawking temperature and entropy of  black holes has been investigated~\cite{Anacleto:2015awa,Anacleto:2015mma,ABCPS2015,ABBS2015,Sakalli:2016mnk,Faizal:2014tea,SG2015,ADV}, 
{showing that in the vicinity of the event horizon the statistical entropy of black holes does not present divergences.}
{In~\cite{KN} has been introduced an expression for the state density modified by the GUP.}
{In~\cite{ABPS,Zhao}, considering the state density modified by the GUP, the statistical entropy was calculated for an acoustic black hole rotating in (2+1) dimensions and the divergences appearing in the brick wall model are eliminated. Thus, no cut-off is required in the state density \cite{Zhang}.}


Since 1981 with Unruh's seminal work~\cite{Unruh,Unruh95}  on models mimicking gravity~\cite{MV, Volovik,Crispino2008,Cadoni2005,MCadoni2005,Andrade2004,Das,Mazur,Lee2001,Stone2004,Oh2005,Hui2012,Barcelo2005,
Berti2004,Cardoso2004,LiChun2011,PAH2004,MAC2013,Vieira,Meng-Sen}, the increase in interest in the study of Hawking radiation has been a very important field of investigation for the understanding of quantum gravity. {In contrast to Hawking radiation, which is a kinematical effect, analog formulae to the Bekenstein-Hawking entropy were unknown for acoustic black holes until recently. However, in Ref.~\cite{Rinaldi} was shown that such analogs indeed arise in a Bose-Einstein condensate (BEC) system as one considers entanglement entropy related to phonons created via the Hawking mechanism - see also \cite{Zhao,ABPS,Meng-Sen}. Interesting enough, this entropy depends on the horizon area of the acoustic black holes and indeed has no relationship with the thermodynamic entropy of the fluid that is zero for a BEC system. Such Bekenstein-Hawking entropy analogue can be computed by using statistical methods. The acoustic black holes entropy is given by the event horizon area with a coefficient that in general depends of the infrared cut-off. Analogously,  as  we shall see shortly, in the present study the acoustic black hole entropy depends on the parameter that controls the GUP. 
As such, this entropy is proportional to the area of the event horizon, but the coefficients are different from 1/4. This is an intrinsic issue that arises in the interpretation of the Bekenstein-Hawking entropy as entanglement entropy even for gravitational black holes \cite{sldkin}.
}

{ In addition,  the authors in~\cite{Xian,Bilic,Liberati,Molina,ABP12,ABP11,ABP10,Anacleto:2013esa} determined the metric of an acoustic black hole that were obtained from relativistic models.}
{In general, a classical treatment as well as a Newtonian is employed in the investigation of analog models of gravitation.}
{ However, in general relativity a black hole is a relativistic gravitational phenomenon.
Thus, a correct treatment  of relativistic system  is obtained if the inertial and gravitational effects of pressure are considered. }
{Nevertheless, the influence of relativistic pressure can be properly incorporated into the dynamics of the Newtonian approach.}
{This change in the structure of the standard Newtonian approach due to the effects of pressure is termed neo-Newtonian theory.}

Once the pressure is generally a parameter that can be easily adjustable and it can {represent the external field source}, 
{thus investigating the analogous Aharonov-Bohm (AB) effect due to a scattering of planar waves by an idealized draining bathtub vortex~\cite{Fetter, Dolan, ABP2012-1,Brito2013,ABPbtz,Anacleto2015lv} is an interesting way of testing analog models via neo-Newtonian theory.}
As shown in~\cite{Anacleto:2015mta}, 
{by analyzing the effect of AB it has been found that the phase change displays a dependence on the pressure and this has analogy with a magnetic flux.}
Furthermore in a recent study \cite{stein} it was observed Hawking radiation by accelerating a fluid due to the pressure of a laser beam. Thus, variation of both Hawking radiation and its entanglement --- which can be measured by entanglement entropy --- seems to be a very important issue in order to control the effect experimentally.
{In~\cite{McCrea} the equations of the neo-Newtonian physics were obtained by McCrea and later in~\cite{Harrison} these equations were improved}.
{In addition, the authors in~\cite{Lima1997} applying a perturbative approach in the neo-Newotinana equations deduced the fluid equations in their final form (see also \cite{RRRR2003, rrrr, ademir, velten, Oliveira}).}
{The authors in \cite{Fabris2013} constructed within  the formalism of the neo-Newtonian hydrodynamics the metric of an acoustic black hole 
and in \cite{Salako:2015tja} was studied the  superresonance phenomenon in the framework of  neo-Newtonian hydrodynamics.}

In this letter, we use the quantum statistical method via corrected state density of the GUP and we calculate the entropy of the rotating acoustic black hole in the neo-Newtonian theory. In our calculations, we have obtained the Bekenstein-Hawking area entropy of acoustic black hole and its correction terms. Again, by considering the GUP on the equation of state density we found that the divergence in the brick-wall model disappears and no cut-off is required. Furthermore, using this method, terms of logarithmic corrections \cite{carlip} are not generated.

\section{Acoustic black holes in neo-Newtonian hydrodynamics}
\label{II}
{In this section we will briefly introduce the metric of an acoustic black hole  in the neo-Newtonian hydrodynamics obtained in \cite{Fabris2013}.
To find the acoustic metric in neo-Newtonian theory we initially redefine the energy density as follows:}
\begin{equation}\label{rhoi}
\rho_i \rightarrow \rho + p.
\end{equation}
{Thus, in the neo-Newtonian scenario the basic equations are~\cite{ademir,rrrr,velten}:}
\begin{equation}
\partial_{t} \rho_i + \nabla\cdot(\rho_i\vec v) + p \nabla\cdot\vec{v} = 0,
\label{salako1bis}
\end{equation}
\begin{equation}\label{salako2}
\dot{ \vec{v}} + (\vec{v} \cdot  \nabla )\vec{v} = -\nabla\Psi- \frac{ \nabla  p}{\rho + p}\,,
\end{equation}
\begin{equation}
\label{salako3}
\nabla^{2} \Psi = 4 \pi G \left(\rho + 3 p\right),
\end{equation}
where $\rho_i$ is the fluid density, $p$ is the pressure and $\vec v$ the {flow/fluid velocity}. The expression (\ref{salako1bis}) is the continuity equation and (\ref{salako2}) is the Euler equation  modified due to gravitational interaction.
In~\cite{McCrea} this result has been generalized in the presence of pressure.
Moreover, in~\cite{Harrison} this approach has been
modified and leads to neo-Newtonian cosmology.
{Note that when $p \ll\rho$ the Newtonian equations are obtained. }


At this point, we consider that the fluid is barotropic, i.e. $ p=p(\rho)$, inviscid and irrotational,  being the equation of state $p = k \rho^{n}$, with $k$ and $n$ constants. We write the fluid velocity as $ \vec{v}=-\nabla\psi $ where $ \psi $ is the velocity potential.
{Then, by neglecting the gravitational effect of Eq. (\ref{salako3}) and applying the usual procedure to obtain the acoustic metric, we linearise the equations (\ref{salako1bis}) and (\ref{salako2})  by introducing perturbations into variables $ \rho $, $ \vec{v} $ and $ \psi $ as follows:}
\begin{eqnarray}
 \rho &=& \rho_{0} + \varepsilon \rho_{1} + {\cal O}(\varepsilon^{2})\;, \\
 \rho^{n} &=& \left[\rho_{0} + \varepsilon \rho_{1}+ {\cal O}(\varepsilon^{2})\right]^{n} \approx
 \rho^{n}_{0} + n \varepsilon \rho^{n-1}_{0} \rho_{1} + ... \;,
 \\
 \vec{v} &=& \vec{v}_{0} + \varepsilon \vec{v}_{1} + {\cal O}(\varepsilon^{2}),
 \\
\psi &=& \psi_{0} + \varepsilon\psi+ {\cal O}(\varepsilon^{2})\;. \label{salako80'}
 \end{eqnarray}
{Then after some algebraic calculations we obtain the wave equation given by}
\begin{eqnarray}
&-&\partial_{t} \Big\{c_{s}^{-2}\rho_{0}\Big[\partial_{t}\psi +
\Big(\frac{1}{2} + \frac{\gamma}{2}\Big)\vec v_{0}.\nabla
\psi\Big]\Big\} +\nabla\cdot\Big\{- c_{s}^{-2}\rho_{0}
\vec{v_{0}}\Big[\Big(\frac{1}{2} +
\frac{\gamma}{2}\Big)\partial_{t}\psi 
+ \gamma\vec{v_{0}}.\nabla \psi\Big]+ \rho_{0}\nabla\psi\Big\}=0,
\label{salako12bis}
\end{eqnarray}
being  $ \gamma=1+kn\rho_0^{n-1} $. The above equation  can still be given as
\begin{eqnarray}
\label{eqf}
\partial_{\mu} (f^{\mu\nu} \partial_{\nu} \psi) = 0.
\end{eqnarray}
{We can also write Eq.~(\ref{eqf}) as a Klein-Gordon equation in the curved space in (2 + 1) dimensions for a massless scalar field given by}
\begin{eqnarray}
 \frac{1}{\sqrt{-g}} \partial_{\mu} ( \sqrt{-g} g^{\mu\nu} \partial_{\nu} \psi) = 0.  \label{salako14'}
\end{eqnarray}
{Thus, by determining the inverse of $ g^{\mu\nu} $, we can write the effective line element in polar coordinates ($ \vec{v}=v_r\hat{r}+v_{\phi}\hat{\phi} $ and $ d\vec{r}=dr\hat{r}+rd\phi\hat{\phi} $) as }
\begin{eqnarray}
\label{salako15''} 
\!\!ds^2\!\!&=&
\!\!\sqrt{\frac{\rho_0}{{\cal C}}}\left[- \left[c^2_s -\gamma (v^2_r+v_{\phi}^2)\right]dt^2-(1+\gamma)(v_r dr+v_{\phi}rd\phi)dt +(dr^2+r^2d\phi^2)
+\frac{(\gamma-1)^2}{4c^2_s}(v_{\phi} dr-v_{r}rd\phi)^2\right],
\end{eqnarray}
{where ${\cal C}=c^2_s+(v^2_r+v_{\phi}^2)\frac{(\gamma-1)^2}{4}$. }
{Now we will introduce the coordinate transformations given by}
\begin{eqnarray}
d\tau=dt +\frac{(1+\gamma)v_r dr}{2(c^{2}_s-\gamma v^2_r)}, \quad\quad
d\varphi=d\phi +\frac{\gamma(1+\gamma)v_r v_{\phi}dr}{r(c^{2}_s-\gamma v^2_r)}.
\end{eqnarray}
{ Hence, we can write the effective line element as}
\begin{eqnarray}
\label{eq-kn} 
 ds^{2} &=& \sqrt{\frac{\rho_0}{{\cal C}}}
 \Bigg \{- \Big[c^2_s- \gamma(v^2_r+v^2_{\phi})  \Big] d\tau^2  + 
 \frac{c^2_s\left[1+(v^2_{r}+v^2_{\phi}) \left(\frac{\gamma -1}{2c_s}\right)^2 \right]}{(c^2_s-\gamma\,v^2_r)} dr^2
 -v_{\phi} (1+\gamma)\, r d\tau d\varphi 
 \nonumber\\
 &+ &
 r^2\Big[1 + v^2_r \left(\frac{\gamma -1}{2c_s}\right)^2\Big] d\varphi^2 \Bigg\}.
\end{eqnarray}
{Notice that as $\gamma-1\ll1$, then $\gamma\sim1$, in the near horizon limit $v_r\sim c_s$ then the diagonal components do not changel, i.e., $g_{\phi\phi}=g_{\varphi\varphi}$ and $g_{rr}$  becomes identical in both equations (\ref{salako15''}) and (\ref{eq-kn}). Since this is the limit we shall adopt throughout the paper the physical results are the same by using (\ref{salako15''}) and (\ref{eq-kn}), though the last equation is easier to deal with.}
Now for a static and position independent density, the {flow/fluid velocity} (which is a solution obtained from the continuity equation (\ref{salako1bis})) is given by
\begin{eqnarray}
\vec{v} = \frac{A}{r} \hat r + \frac{B}{r} \hat\phi,
\label{salako17''}
\end{eqnarray}
and the velocity potential is
\begin{eqnarray}
\label{salako18'}
 \psi(r,\phi) = - A \ln r - B \phi.    
\end{eqnarray}
{Thus, by adopting $ c_s=1 $, substituting (\ref{salako17''})
into the metric (\ref{eq-kn}) and neglecting irrelevant factors independent of position, we obtain the following element of effective line in neo-Newtonian theory~\cite{Anacleto:2015mta} }
\begin{eqnarray}
\label{m-ab-nn}
ds^2&=&\beta_1\left[-\left(1-\frac{\tilde{r}^2_e}{r^2}\right)d\tau^2+(1+\beta_2)\left(1-\frac{\tilde{r}_h^2}{r^2}\right)^{-1} dr^2-{2B\beta_3}d\tau d\varphi+\Big(1+\frac{\beta_4}{r^2} \Big) r^2d\varphi^2\right].
\end{eqnarray}
where 
\begin{eqnarray}\label{betas}
\beta_1 &=&\left(1+\beta_2\right)^{-1/2},
\quad\quad
\beta_2 =\frac{r^2_e}{r^2}\left(\frac{\gamma-1}{2}\right)^2,
\\
\beta_3 &=&\frac{(1+\gamma)}{2}, \quad\quad \beta_4=\left(\frac{A(\gamma-1)}{2}\right)^2,
\end{eqnarray}
{being  $ \tilde{r}_e $ the radius of the ergosphere which satisfies $ g_{\tau\tau}(\tilde{r}_e)=0 $ and in general is larger than $\tilde{r}_h$, the event horizon, i.e.,}
\begin{eqnarray}
\tilde{r}_e=\sqrt{\gamma(A^2+B^2)}=\sqrt{\gamma}r_e,  \quad\quad \tilde{r}_h=\sqrt{\gamma}\vert A\vert=\sqrt{\gamma}r_h.
\end{eqnarray}
{ Therefore, we can write the acoustic black hole metric (\ref{m-ab-nn}) in the form}
\begin{eqnarray}
g_{\mu\nu}=\beta_1\left[\begin{array}{clcl}
-f_1 &\quad\quad\quad 0& -{B\beta_3}\\
0 & \quad (1+\beta_2)f_2^{-1}& 0\\
-{B\beta_3} &\quad\quad\quad 0 & \left( 1+\frac{\beta_4}{r^2}\right)r^2
\end{array}\right],
\end{eqnarray}
where
\begin{eqnarray}
f_1&=&1-\frac{\tilde{r}_e^2}{r^2}, \quad\quad f_2=1-\frac{\tilde{r}_h^2}{r^2}.
\end{eqnarray}
{Now we obtain the Hawking temperature of the acoustic black hole as }
\begin{eqnarray}
\label{temp}
\tilde{T}_{h}&=&\frac{\kappa}{2\pi} =\frac{1}{4\pi}\frac{d(\beta_1 f_2)}{dr}\Big |_{r=\tilde{r}_h}
\\
&=&
\frac{1}{2\pi r_h}\Bigg[\gamma\Bigg(1+\left(\frac{\gamma-1}{2}\right)^2
\Bigg(1+\frac{ B^2}{r_h^2}\Bigg)\Bigg)\Bigg]^{-1/2}.
\end{eqnarray}
Since the equation of state of the present fluid is $ p=k\rho^n $ then $ \gamma=1+kn\rho_0^{n-1}=1+np_0/\rho_0 $. {Thus, for $np_0/\rho_0\ll 1  $ the Hawking temperature can be written  as follows}
\begin{eqnarray}
\label{temp2}
\tilde{T}_{h}&=&\Bigg[1-\frac{np_0}{2\rho_0}+\frac{1}{4}\Big(\frac{np_0}{\rho_0}\Big)^2\Bigg]T_h-\frac{\pi^2B^2}{2}
\Big(\frac{np_0}{\rho_0}\Big)^2T_h^3+\cdots ,
\end{eqnarray}
where $ T_h=\frac{1}{2\pi r_h} $,
while the Unruh temperature for an observer at a distance $r$ is
\begin{eqnarray}
T=\frac{a}{4\pi}&=&\frac{f^{\prime}(\tilde{r}_h)}{4\pi}F^{-1/2}(r).
\end{eqnarray}
They satisfy the following relation
\begin{eqnarray}
\tilde{T}_h=\sqrt{F(r)}T=\frac{f^{\prime}(\tilde{r}_h)}{4\pi}.
\end{eqnarray}
where $f(r)=\beta_1f_2$  and 
\begin{eqnarray} 
 F(r)=-\tilde{g}_{tt}=- \frac{g_{\tau\tau}\,g_{\varphi\varphi}\,-\,g^2_{\tau\varphi}}{g_{\varphi\varphi}}
 = \beta_1 \Bigg(f_1 + \frac{ B^2\,\beta^2_3 }{r^2 + \beta_4 }\Bigg)
 =\beta_1 \Bigg[f_2-\frac{B^2}{r^2}\Bigg(\gamma - \frac{\beta^2_3 }{1 + \beta_4/r^2 }\Bigg)\Bigg].
\end{eqnarray}
{Since we are considering $ \gamma-1=np_0/\rho_0\ll 1  $, we obtain
\begin{eqnarray}
\label{fg}
F(r)=-\tilde{g}_{tt}=\beta_1\left( 1-\frac{\tilde{r}_h^2}{r^2}\right) + {\cal O}\Big(\frac{n^2p_0^2}{\rho_0^2}\Big).
\end{eqnarray}
Thus, for $ r=\tilde{r}_h $ we have $ F(\tilde{r}_h)=-\tilde{g}_{tt}(\tilde{r}_h)=0$. 

Now we can write the line element (\ref{m-ab-nn}) in the diagonal form as follows }
\begin{eqnarray}
\label{md}
ds^2&=&-\left(-\tilde{g}_{tt}\right)d{\tau}^2
+(1+\beta_2)\beta_1^2\left(-\tilde{g}_{tt}\right)^{-1} dr^2+\beta_{1}\Big(1+\frac{\beta_4}{r^2} \Big) r^2d\varphi^2.
\end{eqnarray}

\section{The statistical entropy}
In this section we consider the quantum statistical mechanics via density of states corrected by the GUP to calculate the entropy of a rotating acoustic black hole in neo-Newtonian theory. 
So, we will consider the following partition function for a system of bosons:
\begin{eqnarray}
\ln Z_0=-\sum_i g_i\ln\Big ( 1-e^{-\beta\varepsilon_i}  \Big),
\end{eqnarray}
and the area element with constant time $t$ coordinate is
$ds=2\pi\sqrt{g_{\varphi\varphi} g_{rr}}dr$.
The density state  modified by the GUP is given by
\begin{equation}
g(\omega)=\frac{p^2e^{-\lambda p^2}}{2\pi},
\end{equation}
where $ p^2=p^{i}p_{i} $ {and $ \lambda$ is related to the squared minimum Compton wavelength which is given in Planck length units. This GUP provides corrections to all orders in the
Planck length --- see \cite{KN} for further details.}
Thus, considering the metric (\ref{md}), the partition function of the system is given by
\begin{eqnarray}
\ln Z&=&-\int 2\pi\sqrt{g_{\varphi\varphi} g_{rr}}dr \sum_i g_i\ln\Big ( 1-e^{-\beta\varepsilon_i}  \Big)
\\
&=&-\int\sqrt{g_{\varphi\varphi} g_{rr}}dr \int_0^{\infty} dg \ln\Big ( 1-e^{-\beta\omega_0}  \Big)
\\
&=&-\int\sqrt{g_{\varphi\varphi} g_{rr}}dr \int_0^{\infty} dp \Big(pe^{-\lambda p^2}\Big)\ln\Big ( 1-e^{-\beta\omega_0}  \Big)
\\
&\approx&\int\sqrt{g_{\varphi\varphi} g_{rr}}dr \int^{\infty}_{m\sqrt{-\tilde{g}_{tt}}} 
\frac{\beta_0 e^{-\lambda p^2}p^2 d\omega}{2\Big (e^{\beta\omega_0} -1 \Big)},
\end{eqnarray}
{where in the last step we have made integration by parts. Here $\beta_0^{-1}=\tilde{T}_h$ is the Hawking temperature, $ \beta=\beta_0\sqrt{-\tilde{g}_{tt}}=T^{-1} $ is the inverse of the temperature obtained by an observer at a point $ r $,  $ \omega_0 $ the energy of the particle, $\omega=\omega_0\sqrt{-\tilde{g}_{tt}}$ the energy of the particle obtained by an observer at a point $ r $ and $ -\tilde{g}_{tt} $ given by equation (\ref{fg}). One should recall that $\omega_0=\sqrt{p^2+m^2}$ and $\beta_0$ is constant.}
Next, we can now determine the free energy of the system as follows
\begin{eqnarray}
F=-\frac{1}{\beta_0}\ln Z=\int\sqrt{g_{\varphi\varphi} g_{rr}}dr \int^{\infty}_{m\sqrt{-\tilde{g}_{tt}}} 
\frac{ e^{-\lambda p^2}p^2 d\omega}{2\Big (e^{\beta\omega_0} -1 \Big)}.
\end{eqnarray}
Now we can find the entropy  of the system in the following way
\begin{eqnarray}
S&=&\beta_0^2\frac{\partial F}{\partial\beta_0}
=\beta_0^2\int\sqrt{g_{\varphi\varphi} g_{rr}}dr \int^{\infty}_{m\sqrt{-\tilde{g}_{tt}}} 
\frac{\omega e^{\beta\omega_0} e^{-\lambda p^2}p^2 d\omega}{2\Big (e^{\beta\omega_0} -1 \Big)^2}
\\
\label{entropia}
&=&\frac{1}{2}\int\sqrt{g_{\varphi\varphi} g_{rr}}dr \int^{\infty}_{m\beta} 
\frac{x e^{x} }{(e^{x} -1 )^2}e^{-\lambda\Big(\frac{x^2}{\beta^2}-m^2\Big)}\Big(\frac{x^2}{\beta^2}-m^2 \Big) dx,
\end{eqnarray}
where 
in the equation above we have defined the relationship $ x=\beta\omega_{0}=\beta_0\omega $.
Also, we made use  of the relation among energy, momentum 
and mass $\frac{\omega^2}{-\tilde{g}_{tt}}=\frac{x^2}{\beta^2}=p^2+m^2$, being $m$ the static mass of particles.
Thus, near the horizon, $ \tilde{g}_{tt}(\tilde{r}_h)\rightarrow 0 $, 
{so that the term $\frac{x^2}{\beta^2}-m^2$ approaches $ \frac{x^2}{\beta^2} $,}
we integrate (\ref{entropia}) with respect to $r$
\begin{eqnarray}
S&=&\frac{1}{2}\int\sqrt{g_{\varphi\varphi} g_{rr}}dr \int^{\infty}_{0} 
\frac{x^3 e^{x} }{\beta^2(e^{x} -1 )^2}e^{-\lambda\frac{x^2}{\beta^2}} dx
=\frac{1}{2\beta_0^2} \int^{\infty}_{0} 
\frac{ dx}{4\sinh^2(x/2)}I(x,\epsilon),
\end{eqnarray}
where
\begin{eqnarray}
\label{Ix}
I(x,\epsilon) &=&\int\frac{ \sqrt{g_{\varphi\varphi} g_{rr}}  }{-\tilde{g}_{tt}}\; x^3e^{-\lambda\frac{x^2}{\beta^2}}dr 
\\
&=& \int{\sqrt{ \frac{(1+\beta_2) (1+\beta_4/r^2)}{f_2 } } }
{ \Bigg[f_2-\frac{B^2}{r^2}\Bigg(\gamma - \frac{\beta^2_3 }{1 + \beta_4/r^2 }\Bigg)\Bigg]^{-1} }\; x^3e^{-\lambda\frac{x^2}{F(r)\beta_0^2}}\;rdr.
\end{eqnarray} 
Since we only consider the quantum field near the black hole horizon,	we take $[\tilde{r}_h , \tilde{r}_h	+\epsilon]$	as	the integral interval with respect to $r$, where $\epsilon$ is a positive small constant. 
{Initially, let us consider the particular case  of non-rotating acoustic black hole. This is done by taking $ B=0 $ in metric (\ref{m-ab-nn}). }
So, from equation (\ref{Ix}) when $r \rightarrow \tilde{r}_h$, $f(r)=\beta_1f_2(r)\approx 2\kappa (r-\tilde{r}_h) $, we find
\begin{eqnarray}
I(x,\epsilon)
=\sqrt{\frac{\alpha_ 2}{\alpha_1}}\int_{\tilde{r}_h}^{\tilde{r}_h +\epsilon}
\frac{(r-\tilde{r}_h)+\tilde{r}_h}{[2\kappa(r-\tilde{r}_h)]^{3/2}}{x}^3 e^{-{\lambda} {x}^2/[2\kappa(r-\tilde{r}_h)\beta^2_0]}dr, 
\end{eqnarray}
where $\alpha_1=\sqrt{1+(\gamma-1)^2/4} $,  $\alpha_2=1+(\gamma-1)^2/4\gamma $  
and $\kappa=2\pi\beta_0^{-1}$ is the surface gravity of the acoustic black hole and by variable substitution 
$ t=\frac{{\lambda} x^2}{4\pi(r-\tilde{r}_h)\beta_0} $, we have
\begin{eqnarray}
I(x,\epsilon)
&=&\sqrt{\frac{\alpha_ 2}{\alpha_1}} \int_{\delta}^{\infty}\left[\frac{\beta_0 {x}^4\sqrt{{\lambda}}}{(4\pi)^2}t^{-3/2} 
+\frac{ \tilde{r}_h\beta_0^2 {x}^2}{4\pi\sqrt{{\lambda}}}t^{-1/2}\right]e^{-t}dt
\\
&=&\sqrt{\frac{\alpha_ 2}{\alpha_1}}\left[\frac{\beta_0 {x}^4\sqrt{{\lambda}}}{(4\pi)^2}\Gamma\Big(-\frac{1}{2},\delta\Big)
+\frac{\tilde{r}_h\beta_0^2 {x}^2}{4\pi\sqrt{{\lambda}}}\Gamma\Big(\frac{1}{2},\delta\Big)\right],
\end{eqnarray}
where $ \delta= \frac{{\lambda} {x}^2}{4\pi\beta_0\epsilon}$ and $ \Gamma(z)=\int_{\delta}^{\infty} t^{z-1}e^{t}dt$ is the incomplete Gamma function.

{The constant $\epsilon$ is the analogue of the {\it invariant distance} between brick wall $r_\epsilon$ and horizon $r_h$ given by $\epsilon\sim\sqrt{\beta_0(r_\epsilon-r_h)}$ \cite{thooft,sldkin}. In the present study $\epsilon$  can be obtained by the smallest length given by generalized uncertainty principle~\cite{KN} }
\begin{equation}
\Delta X\Delta P=\frac{1}{2}e^{({\lambda}(\Delta P)^2+\langle P\rangle)^2},
\end{equation}
and the least uncertainty of location 	$\sqrt{e{\lambda}/2}$ can be determinate. Now, if we consider it as a least length of pure space line element, we have
\begin{equation}
\label{elamb}
\sqrt{\frac{e{\lambda}}{2}}=\int_{\tilde{r}_{h}}^{\tilde{r}_h+\epsilon}\sqrt{g_{rr}}dr
\approx\int_{\tilde{r}_{h}}^{\tilde{r}_h+\epsilon}\frac{dr}{\sqrt{2\kappa(r-\tilde{r}_h)}}
=\sqrt{\frac{2\epsilon}{\kappa}}.
\end{equation}
Thus, from (\ref{elamb}), we have $ \delta= \frac{ {x}^2}{2\pi^2 e}$ and {we obtain the entropy
\begin{eqnarray}
S&=&\frac{1}{2\beta_0^2} \sqrt{\frac{\alpha_ 2}{\alpha_1}}\int^{\infty}_{0} 
\frac{ d{x}}{4\sinh^2({x}/2)}\left[\frac{\beta_0 {x}^4\sqrt{{\lambda}}}{(4\pi)^2}\Gamma\Big(-\frac{1}{2},\delta\Big)
+\frac{\tilde{r}_h\beta_0^2 {x}^2}{4\pi\sqrt{{\lambda}}}\Gamma\Big(\frac{1}{2},\delta\Big)\right].
\end{eqnarray}
Now {just rescaling} $ {x}\rightarrow 2{x} $, we have $ \delta= \frac{2 {x}^2}{\pi^2 e}$ such that
\begin{eqnarray}\label{entropy-final}
S&=&\sqrt{\frac{\alpha_ 2}{\alpha_1}}\left[\frac{4\sqrt{{\lambda}}}{(4\pi)^2\beta_0}\delta_1
+\frac{\tilde{r}_h}{4\pi\sqrt{{\lambda}}}\delta_2\right],
\end{eqnarray}
being
 \begin{eqnarray}
\delta_1=\int^{\infty}_{0} 
\frac{{x}^4}{\sinh^2({x})}\Gamma\Big(-\frac{1}{2},\delta\Big)d{x}, \quad\quad 
\delta_2=\int^{\infty}_{0}\frac{{x}^2}{\sinh^2({x})}\Gamma\Big(\frac{1}{2},\delta\Big)d{x}.
\end{eqnarray}
{Finally, assuming that $ \sqrt{{\lambda}}=\frac{\delta_2}{2\pi^2} $ (in units of Planck length) and since $ \delta_2=2.017 $ we have obtained $ \lambda=0.0145 $ for the GUP parameter. This is in fairly good agreement with the lower bound obtained in four dimensions $\lambda\gtrsim0.0187$ found in \cite{Abbasvandi:2016oyw} --- see also \cite{Brau:1999uv}.}
Thus, for entropy we find 
\begin{eqnarray}
S=\frac{1}{4}\sqrt{\frac{\gamma\alpha_ 2}{\alpha_1}}(2\pi r_h)
+\sqrt{\frac{\alpha_ 2}{\alpha_1}}\frac{\delta_1\delta_2}{8\pi^4}\tilde{T}_h,
\end{eqnarray}
where $ 2\pi r_h $ is the horizon area of the acoustic black hole. {One should mention that this choice aims to keep the coefficient 1/4  of area in Bekenstein-Hawking entropy analogue and also to put some bound on the parameter $\lambda$. On the other hand, even though we leave the parameter $\lambda$ free, the formula (\ref{entropy-final})  reproduces the acoustic black hole entropy in terms of the event horizon area up to correcting terms.  Furthermore, as we have anticipated in the introduction, the pre-factor 1/4 is not necessarily obtained in the statistical description of the Bekenstein-Hawking entropy as an entanglement entropy --- see further discussions on this issue in Ref.~\cite{sldkin}. 

} The second term is a correction term to the area entropy and is proportional to the radiation temperature, 
\begin{eqnarray}
\tilde{T}_{h}=
\Bigg[\gamma+\frac{\gamma(\gamma-1)^2}{4}\Bigg]^{-1/2}T_h= \Bigg(1- \frac{np_0}{2\rho_0} +\cdots \Bigg)T_h,
\end{eqnarray}
of the acoustic black hole. For $ \frac{np_0}{\rho_0}\ll 1 $ the entropy becomes
\begin{eqnarray}
S=\frac{1}{4}\Bigg(1+ \frac{np_0}{2\rho_0} +\cdots \Bigg)(2\pi r_h)
+\Bigg(1-\frac{np_0}{2\rho_0}+\cdots\Bigg)\frac{\delta_1\delta_2}{8\pi^4}{T}_h.
\end{eqnarray}
{Now, for the case $ B\neq 0 $ {and near the horizon we approach} $ f_2\Bigg[f_2-\frac{B^2}{r^2}\Bigg(\gamma - \frac{\beta^2_3 }{1 + \beta_4/r^2 }\Bigg)\Bigg]^2 \approx [2\kappa (r-\tilde{r}_h)]^3$, so  we have formally the same equation for the entropy
\begin{eqnarray}
S=\frac{1}{4}\sqrt{\frac{\gamma\alpha_ 2}{\alpha_1}}(2\pi r_h)
+\sqrt{\frac{\alpha_ 2}{\alpha_1}}\frac{\delta_1\delta_2}{8\pi^4}\tilde{T}_h,
\end{eqnarray}
with the modified constant $\alpha_1$ given in terms of the parameter $B$}
\begin{eqnarray}
\alpha_1=\sqrt{1+\frac{(\gamma-1)^2}{4}\left( 1+\frac{\gamma B^2}{r_h^2} \right)},
\end{eqnarray}
so that for $ \frac{np_0}{\rho_0}\ll 1 $ we obtain
\begin{eqnarray}\label{entropy-b}
S&=&\frac{1}{4}\Bigg(1+ \frac{np_0}{2\rho_0} -\frac{(np_0)^2}{16\rho_0^2} \Bigg)(2\pi r_h)
-\frac{\pi^2B^2}{16}\frac{(np_0)^2}{\rho_0^2}T_h
+\Bigg(1-\frac{np_0}{2\rho_0}+\frac{5}{16}\frac{(np_0)^2}{\rho_0^2} \Bigg)\frac{\delta_1\delta_2}{8\pi^4}{T}_h
\nonumber\\
&-&\frac{3B^2\delta_1\delta_2}{32\pi^2}\frac{(np_0)^2}{\rho_0^2} T_h^3+\cdots.
\end{eqnarray}
Note that for the first term we have obtained corrections to the area of the horizon of the acoustic black hole in neo-Newtonian theory. {The following terms are dependent on the radiation temperature of the acoustic black hole. Namely, the second term is a correction term that arises from the contribution  due to the  presence of pressure and parameter $B$ associated with the rotation of the acoustic black hole. The third term is a correction term that depends on the pressure and thermal radiation. The fourth term is a correction to the entropy that  depends on the rotation parameter and pressure. In addition these last terms are due to the GUP since they depend on the parameters $\delta_1, \delta_2$.}
Moreover, the calculation of entropy for this method does not generate logarithmic corrections {due to the GUP (corrections of these types in 2+1 dimensions for BTZ black holes have been well addressed in Ref.~\cite{carlip} and in Ref.~\cite{Anacleto:2015awa} the application of the GUP to a specific planar black hole solution revealed logarithm corrections)}. 

In a closely related study we could also identify the Aharonov-Bohm phase shift  depending on the pressure is analogous to a magnetic flux \cite{Anacleto:2015mta}. A related reasoning has been considered in graphene physics \cite{43}.

\section{Conclusions}

In summary,  in the present study we calculate the entropy of an acoustic black hole in the context of neo-Newtonian hydrodynamics. Applying quantum statistical method we obtain corrections to the entropy due to the GUP. 
The use of the GUP in the equation of state density allows us {to compute the partition function} without the requirement of a cut-off and the divergences that arise when we apply the brick-wall method are eliminated.
In our results we found that the leading term in Eq. (\ref{entropy-b}) is proportional to  horizon area of the acoustic black hole and terms of corrections are proportional to $ T_h $ and $T_h^3$. {Similar results were found in acoustic black holes in 2+1 dimensions \cite{Zhao,ABPS} and BTZ black holes in \cite{ABCPS2015}. The computation of corrected entropy with this GUP for Schwarzschild black hole in 3+1 was first computed in \cite{KN}.} In addition, we have obtained a new term correction that is proportional to $T_h ^ 3$ that arises due to the presence of pressure and $B$ parameter associated with the rotation of the acoustic black hole. Since the pressure is generally a parameter that can be easily adjustable and it can play the role of an external field, the neo-Newtonian theory might provide an interesting way to test analog effects, possibly in some variants of the recent experiment developed in Ref.\cite{stein}. 



\acknowledgments

We would like to thank CNPq and CAPES for partial financial support. 
Ines G. Salako thanks  IMSP  for hospitality during the elaboration of this
work.


\begin{thebibliography}{100}
\bibitem{Hawking} S. W. Hawking, Phys. Rev. Lett. {\bf 25},1344 (1971).
 
\bibitem{SWH} S. W. Hawking, Commun. Math. Phys. {\bf 43}, 199 (1975). 

\bibitem{JDB} J. D. Bekenstein, Phys. Rev. D {\bf 7},  2333 (1973).

\bibitem{JDB85} J. D. Bekenstein, Phys. Rev. D {\bf 9}, 3292 (1974).

\bibitem{SWU} M. Srednicki, Phys. Rev. Lett. {\bf71} (1993) 666;
C. G. Callan, Jr., F. Wilczek, Phys. Lett. B{\bf333} (1994)
55; L. Susskind, J. Uglum, Phys. Rev. D {\bf50} (1994)
2700.

\bibitem{G-H}G. W. Gibbons, S. W. Hawking, Phys. Rev. D {\bf15} (1977)
2752.

\bibitem{thooft} G. `t Hooft, Nucl. Phys. B {\bf 256}, 727 (1985).

\bibitem{Rinaldi} M. Rinaldi, Phys. Rev. D {\bf 84}, 124009 (2011).

\bibitem{Rinaldi:2013aa} M.~Rinaldi, 
Int.\ J.\ Mod.\ Phys.\ D {\bf 22} (2013) 1350016, [arXiv:1112.3596 [gr-qc]].  
   
\bibitem{SG2011} S. Giovanazzi, Phys. Rev. Lett. 106, 011302 (2011).   
  
\bibitem{Brustein}   Ram Brustein, and Judy Kupferman, 
Phys. Rev. D {\bf 83}, 124014 (2011), [arXiv:1010.4157 [hep-th]]. 

\bibitem{Kim2006} Kim, Wontae, Kim, Yong-Wan, and Park, Young-Jai, 
J. Korean Phys. Soc., 49, 1360, (2006), [arXiv:gr-qc/0604065 [gr-qc]].
 
\bibitem{Park2007} Kim, Yong-Wan, and Park, Young-Jai, 
Phys. Lett., B {\bf 655}, 172 (2007), [arXiv:0707.2128 [gr-qc]]. 

\bibitem{Sun2004} Sun, Xue-Feng, and Liu, Wen-Biao, 
Mod. Phys. Lett., A {\bf 19}, 677 (2004).
 
\bibitem{Yoon2007} Yoon, Myungseok, Ha, Jihye, and Kim, Wontae, 
Phys. Rev., D {\bf 76}, 047501 (2007), 
[arXiv:0706.0364 [gr-qc]].

\bibitem{XLi} X. Li, Z. Zhao, Phys. Rev. D {\bf 62}, 104001 (2000).

\bibitem{zhao2004} R. Zhao, S. L. Zhang, Gen. Relat. Grav. {\bf 36}, 2123 (2004).

\bibitem{RZ2003} R. Zhao, Y. Q. Wu, L. C. Zhang, Class. Quantum Grav. {\bf 20}, 4885 (2003).

\bibitem{WK2006} W. Kim, Y. W. Kim, Y. J. Park, Phys. Rev. D {\bf 74}, 104001 (2006).

\bibitem{KKP2007} W. Kim, Y. W. Kim, Y. J. Park, Phys. Rev. D {\bf 75}, 127501 (2007).

\bibitem{JHa2007} M. Yoon, J. Ha, W. Kim, Phys. Rev. D {\bf 76},  047501 (2007).

\bibitem{YWKim2007} Y. W. Kim, Y. J. Park, Phys. Lett. B {\bf 655}, 172 (2007).

\bibitem{Anacleto:2015awa} 
  M.~A.~Anacleto, F.~A.~Brito, G.~C.~Luna, E.~Passos and J.~Spinelly,
  Annals Phys.\  {\bf 362}, 436 (2015)
  doi:10.1016/j.aop.2015.08.009
  [arXiv:1502.00179 [hep-th]].
  
\bibitem{Anacleto:2015mma} 
  M.~A.~Anacleto, F.~A.~Brito and E.~Passos,
  Phys.\ Lett.\ B {\bf 749}, 181 (2015)
  doi:10.1016/j.physletb.2015.07.072
  [arXiv:1504.06295 [hep-th]].
   
  
\bibitem{ABCPS2015} 
  M.~A.~Anacleto, F.~A.~Brito, A.~G.~Cavalcanti, E.~Passos and J.~Spinelly,
  Gen.\ Rel.\ Grav.\  {\bf 50}, no. 2, 23 (2018)
  doi:10.1007/s10714-018-2344-x
  [arXiv:1510.08444 [hep-th]].
  

\bibitem{ABBS2015}  
M.~A.~Anacleto, D.~Bazeia, F.~A.~Brito and J.~C.~Mota-Silva,
  arXiv:1512.07886 [hep-th].  
  
\bibitem{Sakalli:2016mnk} 
  I.~Sakalli, A.~Ovgun and K.~Jusufi,
  arXiv:1602.04304 [gr-qc].
  
  \bibitem{Faizal:2014tea} 
  M.~Faizal and M.~M.~Khalil,
  Int.\ J.\ Mod.\ Phys.\ A {\bf 30}, no. 22, 1550144 (2015)
  [arXiv:1411.4042 [gr-qc]].

\bibitem{SG2015}  
   S. Gangopadhyay, A. Dutta and M. Faizal, Europhys. Lett. {\bf 112}, no. 2, 20006 (2015) doi:10.1209/0295-5075/112/20006
[arXiv:1501.01482 [gr-qc]].
  
\bibitem{ADV} A. F. Ali, S. Das and E. C. Vagenas, Phys. Lett. B {\bf 678}, 497 (2009).

\bibitem{KN} K. Nouicer, Phys. Lett. B {\bf 646}, 63 (2007).

\bibitem{ABPS} 
  M.~A.~Anacleto, F.~A.~Brito, E.~Passos and W.~P.~Santos,
  Phys.\ Lett.\ B {\bf 737}, 6 (2014)
  [arXiv:1405.2046 [hep-th]].
\bibitem{Zhao} HuiHua Zhao, GuangLiang Li, LiChun Zhang, Phys.Lett. A {\bf 376},  2348 (2012).

\bibitem {Zhang} R. Zhao, L.C. Zhang, H.F. Li, Acta Phys. Sin. 58,  2193 (2009).

\bibitem{Unruh} W. Unruh, Phys. Rev. Lett. {\bf 46}, 1351 (1981).

\bibitem{Unruh95} W. Unruh, Phys. Rev. D {\bf 51}, 2827 (1995), [arXiv:gr-qc/9409008]. 

\bibitem{MV} M. Visser, Class. Quant. Grav. {\bf 15} 1767 (1998).

\bibitem{Volovik} G. Volovik, {\it The Universe in a Helium Droplet}, Oxford University Press, 2003.

\bibitem{Crispino2008} L. C. B. Crispino, A. Higuchi, and G. E. A. Matsas, Rev. Mod. Phys. {\bf 80}, 787 (2008), [arXiv:0710.5373[gr-qc]].

\bibitem{Cadoni2005} M. Cadoni, S. Mignemi, Phys. Rev. D {\bf 72}, 084012 (2005).

\bibitem{MCadoni2005} M. Cadoni, Class. Quant. Grav. {\bf 22}, 409 (2005).

\bibitem{Andrade2004} L. C. Garcia de Andrade, Phys. Rev. D {\bf 70} (2004) 64004.

\bibitem{Das} T. K. Das, {\it Transonic black hole accretion as analogue system}, [arXiv:gr-qc/0411006].

\bibitem{Mazur} G. Chapline, P.O. Mazur, {\it Superfluid picture for rotating space–times},[arXiv:gr-qc/0407033].

\bibitem{Lee2001} O. K. Pashaev, J.-H. Lee, Theor. Math. Phys. 127 (2001) 779, [arXiv:hep-th/0104258].

\bibitem{Stone2004} S. E. Perez Bergliaffa, K. Hibberd, M. Stone, M. Visser, Physica D {\bf 191} (2004) 121,
[arXiv:cond-mat/0106255].

\bibitem{Oh2005} S. W. Kim, W. T. Kim, J. J. Oh, Phys. Lett. B {\bf 608} (2005) 10, [arXiv:gr-qc/0409003].

\bibitem{Hui2012} Xian-Hui Ge, Shao-Feng Wu, Yunping Wang, Guo-Hong Yang, You-Gen Shen,  Int. J. Mod. Phys. {\bf D21} (2012) 1250038, 
[arXiv:1010.4961 [gr-qc]].

\bibitem{Barcelo2005} C. Barcelo, S. Liberati, M. Visser, Living Rev. Rel. 8 (2005) 12, [arXiv:gr-qc/0505065].

\bibitem{Berti2004} E. Berti, V. Cardoso, J.P.S. Lemos, Phys. Rev. D {\bf 70} (2004) 124006, [arXiv:gr-qc/0408099].

\bibitem{Cardoso2004} V. Cardoso, J.P.S. Lemos, S. Yoshida, Phys. Rev. D {\bf 70} (2004) 124032, [arXiv:gr-qc/0410107].

\bibitem{LiChun2011} Li-Chun Zhang, Huai-Fan Li, Ren Zhao, Phys. Lett. B {\bf 698} (2011) 438.
 
\bibitem{PAH2004} Peter A. Horvathy, Mikhail S. Plyushchay,
Phys. Lett. B {\bf 595} (2004) 547, [hep-th/0404137].
 
\bibitem{MAC2013} M.~A.~Cuyubamba,
  Class.\ Quant.\ Grav.\  {\bf 30}, 195005 (2013)
  [arXiv:1304.3495 [gr-qc]].
   
\bibitem{Vieira}  H.~S.~Vieira and V.~B.~Bezerra,
  arXiv:1406.6884 [gr-qc].

\bibitem{Meng-Sen} Ma Meng-Sen et al, Commun. Theor. Phys. {\bf60} (2013) 695Ð698.

\bibitem{sldkin} S. N. Solodukhin, {\it Entanglement entropy of black holes},
[arXiv:1104.3712 [hep-th]].

\bibitem{Xian} X.-H. Ge, S.-J. Sin, JHEP {\bf 1006}, 087 (2010) [arXiv:hep-th/1001.0371].
 
\bibitem{Bilic} N. Bilic, Class. Quant. Grav. {\bf16} (1999) 3953.
 
\bibitem{Liberati} S. Fagnocchi, S. Finazzi, S. Liberati, M. Kormos and A. Trombettoni [arXiv:1001.1044[gr-qc]].

\bibitem{Molina} M. Visser and C. Molina-Paris, [arXiv:1001.1310[hep-th]].

\bibitem{ABP12} M. A. Anacleto, F. A. Brito, E. Passos, Phys. Rev. D {\bf 85}, 025013 (2012)  [arXiv:1109.6298 [hep-th]].

\bibitem{ABP11} M. A. Anacleto, F. A. Brito, E. Passos, Phys. Lett. B {\bf703}, 609 (2011)  [arXiv:1101.2891 [hep-th]].

\bibitem{ABP10} M. A. Anacleto, F. A. Brito, E. Passos Phys. Lett. B {\bf 694}, 149 (2010)  [arXiv:1004.5360 [hep-th]].

\bibitem{Anacleto:2013esa} 
  M.~A.~Anacleto, F.~A.~Brito and E.~Passos,
  Phys.\ Lett.\ A {\bf 380}, 1105 (2016)
  doi:10.1016/j.physleta.2016.01.030
  [arXiv:1309.1486 [hep-th]].


\bibitem{Fetter} A. L. Fetter, Phys. Rev. 136 (1964) A1488. 

\bibitem{Dolan}S.R. Dolan, E.S. Oliveira, L.C.B. Crispino, Phys. Lett. B {\bf 701}, 485 (2011).

\bibitem{ABP2012-1} M.~A.~Anacleto, F.~A.~Brito and E.~Passos, Phys. Rev. D {\bf 86}, 125015 (2012) 
[arXiv:1208.2615 [hep-th]]; 

\bibitem{Brito2013} M.~A.~Anacleto, F.~A.~Brito and E.~Passos Phys. Rev. D {\bf 87}, 125015 (2013) [arXiv:1210.7739 [hep-th]].

\bibitem{ABPbtz} M.~A.~Anacleto, F.~A.~Brito and E.~Passos,
  Phys.\ Lett.\ B {\bf 743}, 184 (2015)
  [arXiv:1408.4481 [hep-th]].
  
\bibitem{Anacleto2015lv}  M.~A.~Anacleto,
  Phys.\ Rev.\ D {\bf 92}, no. 8, 085035 (2015)
  doi:10.1103/PhysRevD.92.085035
 [ arXiv:1505.03238 [hep-th]].
  
\bibitem{Anacleto:2015mta} 
  M.~A.~Anacleto, I.~G.~Salako, F.~A.~Brito and E.~Passos,
  Phys.\ Rev.\ D {\bf 92}, no. 12, 125010 (2015)
  doi:10.1103/PhysRevD.92.125010
  [arXiv:1506.03440 [hep-th]]. 
  
  \bibitem{stein} 
  J.~Steinhauer, Nature Physics {\bf12}, 959Ð965 (2016),
  doi:10.1038/nphys3863
  [arXiv:1510.00621 [gr-qc]].
    

\bibitem{McCrea} W. H. McCrea, Proc. R. Soc. London {\bf206}, 562 (1951).

\bibitem{Harrison} E. R. Harrison, Ann. Phys (N.Y.) {\bf35}, 437 (1965).

\bibitem{Lima1997} J. A. S. Lima, V. Zanchin and R. Brandenberger, Month. Not. R. Astron. Soc. {\bf291}, L1 (1997).

\bibitem{RRRR2003} R. R. R. Reis, Phys. Rev. {\bf D67}, 087301 (2003); erratum-ibid {\bf D68}, 089901(2003).

\bibitem{ademir} J .A. S. Lima, V. Zanchin and R. Brandenberger, MNRAS, {\bf 291}, L1-L4 (1997).

\bibitem{rrrr} R. R. R. Reis, Phys. Rev. {\bf D67}, 087301 (2003); erratum-ibid {\bf D68}, 089901 (2003).

\bibitem{velten} H. Velten, D. J. Schwarz, J. C. Fabris and W. Zimdahl, Phys Rev D {\bf 88}, 103522 (2013).

\bibitem{Oliveira} A. M. Oliveira, H. E. S. Velten, J. C. Fabris, I. G. Salako, Eur. Phys. J. C {\bf 74} 3170 (2014).

\bibitem{Fabris2013}
J. C. Fabris, O. F. Piattella, I. G. Salako, J. Tossa, H. E. S. Velten, Mod. Phys. Lett A {\bf28}, 1350169 (2013), 
[arXiv:1308.1859 [gr-qc]].


\bibitem{Salako:2015tja}
I.~G.~Salako and A.~Jawad,
Int.\ J.\ Mod.\ Phys.\ D {\bf 25}, no. 05, 1650055 (2016), doi:10.1142/S0218271816500553
[arXiv:1503.08714 [gr-qc]].

\bibitem{carlip} S. Carlip, Class. Quant. Grav. {\bf17}, 4175 (2000) [gr-qc/0005017].

\bibitem{Abbasvandi:2016oyw} 
  N.~Abbasvandi, M.~J.~Soleimani, S.~Radiman and W.~A.~T.~Wan Abdullah,
  Int.\ J.\ Mod.\ Phys.\ A {\bf 31}, no. 23, 1650129 (2016)
  doi:10.1142/S0217751X16501293
  [arXiv:1607.05863 [hep-th]].
  
\bibitem{Brau:1999uv} 
  F.~Brau,
  J.\ Phys.\ A {\bf 32}, 7691 (1999)
  doi:10.1088/0305-4470/32/44/308
  [quant-ph/9905033].

\bibitem{43} F. de Juan, A. Cortijo, M. A. H. Vozmediano and A. Cano, Nature Physics {\bf7}, 811 (2011), [arXiv:1105.0599 [cond-mat.other]].  

\end{thebibliography}
\end{document}